# Expression and interactions of stereo-chemically active lone pairs and their relation to structural distortions and thermal conductivity

Kasper Tolborg,[a] Carlo Gatti,[b] Bo B. Iversen[a]*

a) Center for Materials Crystallography, Department of Chemistry and iNANO, Aarhus University, Langelandsgade 140, 8000 Aarhus C, Denmark

b) CNR-SCITEC Istituto di Scienze e Tecnologie Chimiche "Giulio Natta", via Golgi section, via Golgi 19, 20133 Milano, Italy

*Corresponding author: Bo B. Iversen, bo@chem.au.dk

## Abstract

In chemistry, stereo-chemically active lone pairs are typically described as an important non-bonding effect, and large recent interest has centered on understanding the derived effect of lone pair expression on physical properties such as the thermal conductivity. To manipulate such properties, it is essential to understand the conditions that lead to lone pair expression and to provide a quantitative chemical description of their identity to allow comparison between systems. Here we first use density functional theory calculations to establish the presence of stereo-chemically active lone pairs on antimony in the archetypical chalcogenide $MnSb_2O_4$. The lone pairs are formed through a similar mechanism to those in binary post-transition metal compounds in an oxidation state of two less than their main group number (e.g. Pb(II) and Sb(III)), where the degree of orbital interaction (covalency) determines the expression of the lone pair. In $MnSb_2O_4$ the Sb lone pairs interact through a void space in the crystal structure, and they minimize their mutual repulsion by introducing a deflection angle. This angle increases significantly with decreasing Sb-Sb distance introduced by simulating high pressure, thus showing the highly destabilizing nature of the lone pair interactions. Analysis of the chemical bonding in $MnSb_2O_4$ shows that it is dominated by polar covalent interactions with significant contributions both from charge accumulation in the bonding regions and from charge transfer. A database search of related ternary chalcogenide structures shows that for structures with a lone pair ($SbX_3$ units) the degree of lone pair expression is largely determined by whether the antimony-chalcogen units are connected or not, suggesting a cooperative effect. Isolated $SbX_3$ units have larger X-Sb-X bond angles, and therefore weaker lone pair expression than connected units. Since increased lone pair expression is equivalent to an increased orbital interaction (covalent bonding), which typically leads to increased heat conduction, this can explain the previously established correlation between larger bond angles and lower thermal conductivity. Thus, it appears that for these chalcogenides, lone pair expression and thermal conductivity may be related through the effects of atomic displacement on the electronic structure, rather than anharmonicity because of specific soft potential directions.

## Introduction

Stereo-chemically active lone pairs are usually treated as text book examples of non-bonding effects and occur in post-transition metal compounds in which the post-transition metal is in an oxidation state of two lower than its main group number, such as Pb(II) and Sb(III). This means that the two outermost s-electrons are available to form a lone pair, leading, after a proper hybridization, to the possibility of an asymmetric coordination environment. These oxidation states are found in many technologically important materials, such as thermoelectric materials,[1, 2] multiferroics,[3] phase-change materials[4] and optoelectronics,[5, 6] but their structural chemistry is quite intriguing, since the ability to form a lone pair does not always lead to asymmetric coordination, e.g. in symmetric rock salt PbS and asymmetric litharge PbO.[7] Originally, and still in most textbooks, the stereo-chemically active lone pairs are described as an on-site sp-hybridization on the metal atom, which is the origin of its name as a *chemically inactive*, i.e. non-bonding, but *stereo-chemically active*, i.e. structure determining, effect. However, this does not properly explain the anion dependence on the tendency to form an asymmetric coordination environment.

In a series of articles, Walsh, Watson and co-workers introduced a revised model to account for this anion dependence based also on work by Waghmare and co-workers.[7-10] They showed that the lone pairs are in fact not chemically inactive, but rather that the s-orbital on the metal atom hybridizes with the anion p-orbital to form a bonding and an anti-bonding state, followed by hybridization of the anti-bonding state with the metal p-states. The tendency to form a lone pair is therefore highly dependent on the energy difference between the metal valence states and the anion valence p-states. If this difference is large, the formation of the bonding orbital is less favorable, and thus the lone pair is not "expressed" and the higher symmetry structure is adopted.[8] Thus, the expression of a lone pair is highly dependent on covalent interactions between cation and anion.

With these trends established, we have a framework for understanding the structures of materials with the possibility of forming stereo-chemically active lone pairs. The relation to physical properties has been investigated for thermoelectric materials, where the presence of oxidation states, which are able to form stereo-chemically active lone pairs, was shown to lead to a decrease in lattice thermal conductivity in the Cu-Sb-Se ternary system, i.e. $Sb^{3+}$ systems have lower thermal conductivity than $Sb^{5+}$ systems.[11] Furthermore, an empirical correlation between bond angle and lattice thermal conductivity in As, Sb and Bi chalcogenides in oxidation state +III was established with larger bond angles generally leading to lower thermal conductivity. The ability to form the lone pair has also been used to describe the good electronic properties for thermoelectricity of rock salt lead and tin monochalcogenides compared with other rock salt materials.[12]

However, the general relation to physical properties and the crystal structures beyond the asymmetric local coordination environment is still poorly understood. The structural point is illustrated by the large difference in crystal structures adopted by otherwise similar materials, where the local coordination environment is asymmetric, such as the localized molecular units in $Sb_2O_3$ and the infinite chains in $Sb_2S_3$. Similarly, the bond angles around the stereo-chemically active lone pair can be very different for otherwise similar systems.[11, 13]

An interesting chalcogenide material in which the local coordination environment is asymmetric is the isostructural group of ternary oxides $MSb_2O_4$, where M is a transition metal. The structure has been reported for M = Mn, Fe, Co, Ni and Zn crystallizing in the space group $P4_2/mbc$, isostructural to red lead, $Pb_3O_4$, which can formally be written as $Pb(IV)Pb(II)_2O_4$ to highlight the similarity between the groups. The crystal structure of $MnSb_2O_4$ is shown in Figure 1 along the *c*- and *a*-axes. The structure consists of distorted $MnO_6$ octahedra with two long (Mn-O1) and four short (Mn-O2) distances, and $SbO_3$ units in a trigonal pyramidal coordination with two long (Sb-O1) and one short (Sb-O2) distances, leaving room for a presumed lone pair on antimony to occupy the fourth corner in a tetrahedron.[14, 15]

The $MSb_2O_4$ group of compounds is relatively unexplored in the literature, but they have been studied for potential use in Li-ion batteries due to presence of channels of low electron density along the *c*-axis, and the magnetic structures have been studied and were shown to result in different orderings depending on the metal atom.[15-18] Our present interest in this class of materials arises from the interesting structure, where the presumed position of the antimony lone pair based on the coordination environment suggests that two lone pairs point almost directly towards each other. Basic chemical intuition would suggest that this interaction should be highly unfavorable, and similar structural motifs are not, to our knowledge, adopted by other materials with stereo-chemically active lone pairs.

Here we investigate the interactions between the presumed stereo-chemically active lone pairs on antimony in $MnSb_2O_4$ through density functional theory (DFT) calculations based on both orbital and electron density related descriptors to gain further understanding of this intriguing structural motif. Furthermore, we simulate the high-pressure behavior of the material to force the lone pairs closer together and investigate the response of this external stimulus. Then, we report a characterization of all other bonding interactions in the material, and finally, we analyze a set of related structures in order to derive general features regarding the influence of stereo-chemically active lone pairs on crystal structures, and their relation to physical properties especially thermal conductivity.

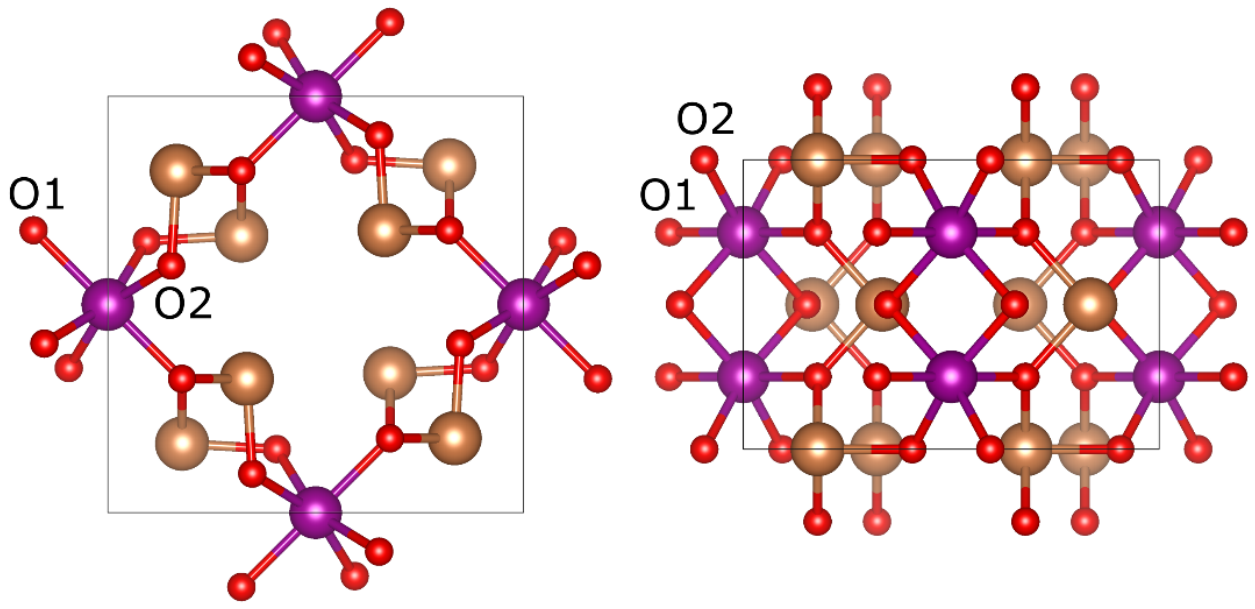

**Figure 1.** Crystal structure of $MnSb_2O_4$ viewed along the *c*-axis (left) and *a*-axis (right). Manganese is shown as purple, antimony as brown and oxygen as red. The two crystallographically unique oxygen atoms are marked as O1 and O2.

## Computational details

Periodic ab initio DFT calculations on $MnSb_2O_4$ were performed in CRYSTAL14 using the POB-TZVP basis set.[19-21] This basis set uses a full-potential all-electron basis for Mn and O and a small-core effective core potential for Sb with 23 electrons in the valence corresponding to the $4s^2 4p^6 4d^{10} 5s^2 5p^3$ electrons. The PBE0 hybrid functional was used and reciprocal space was sampled on an 8x8x8 grid in a Monkhorst-Pack net in the first Brillouin zone.[22, 23] The initial geometry and magnetic symmetry was set to the geometry from Roelsgaard *et al.*,[15] and the antiferromagnetic configuration corresponding to the A-mode from Fjellvåg and Kjekshus.[16] This in principle results in lowering the space group symmetry to $P\bar{4}b2$, which would require displacement of all atomic z-coordinates with ¼ to follow the standard settings of the space groups. However, in this case it was done by using the CRYSTAL14 keyword MODISYMM to remove (half of) the symmetry elements to still have the coordinates corresponding to the ones in the structural space group. A ferromagnetic configuration was also tried, but resulted in higher energy and similar bonding features.

First, the full geometry, i.e. cell and atomic coordinates, was optimized with the symmetry from the magnetic structure. The resulting atomic coordinates still followed the structural space group symmetry within the numerical error. Results of the optimization are given in Supporting Note 1 in the Supporting Information (SI). Here it is seen that the deviation of the optimized geometry from the experimental structure at 100 K is less than 0.5 % for cell parameters and less than 2 % for bond lengths. After this, a series of unit cell volumes from 10 % smaller to 4 % larger than the equilibrium volume in steps of 2 % were constructed and the cell parameters and atomic coordinates were relaxed at constant volume (CVOLOPT keyword). The energy-volume curve was fitted to a third-order Birch-Murnaghan equation-of-state to find the corresponding pressures.

The chemical bonding was analyzed in terms of projected density of states and valence electron density as implemented in CRYSTAL14, and topological analysis following Bader's Quantum Theory of Atoms in Molecules (QTAIM)[24] using TOPOND interfaced with CRYSTAL14.[25]

## Results & Discussion

### Density of states and valence electron density

To establish the presence of a stereo-chemically active lone pair on antimony, we first investigate the orbital projected density of states (DOS) and the valence electron density in the region of interest. The DOS (Figure 2) can be divided into four distinct regions in accordance with previously established stereo-chemically active lone pairs in $Sb_2O_3$ and tin monochalcogenides.[9, 26] Region IV is special for this material compared with e.g. $Sb_2O_3$ due to the transition metal and consists mainly of manganese d-states and oxygen p-states. A similar feature is also seen in region III, but here we also observe a large degree of overlap between oxygen p-states and antimony s- and p-states. Further below the Fermi level, we have region II, which consists mainly of oxygen p-states and antimony p-states, and region I, which consists mainly of antimony s-states and oxygen p-states. Regions I, II and III are qualitatively very similar to the ones in e.g. $Sb_2O_3$ with the complication of the presence of a transition metal here.[26]

To further highlight the character of these regions, we can plot the valence electron density within these energy intervals. The most interesting regions regarding the stereo-chemically active lone pair are regions I and III, which are shown in Figure 3. In Figures 3a and 3b, we see that region I consists of a bonding interaction between Sb and O, both to O1 and O2. However, we clearly observe that the electron density in this energy range is higher for the long Sb-O1 bond. This oxygen is bonded to two Sb and one Mn, which seems to affect the energy levels of its valence states. This indicates that the interaction between Sb and O1 is the dominating feature for the expression of the lone pair.

In region III (Figures 3c and 3d), the lone pair on Sb and localized electron density on O are clearly observed. This region is identified as the key for the stereo-chemically active lone pair, and this valence density is in very good correspondence with the density observed in e.g. SnO by Walsh and Watson,[9] which is one of the archetypical examples of stereo-chemically active lone pairs induced by the anion. Thus, it is safe to conclude that the present material has stereo-chemically active lone pairs on antimony following the established framework; however, with some complexity induced by the presence of the transition metal making e.g. the different oxygen atoms behave quite differently.

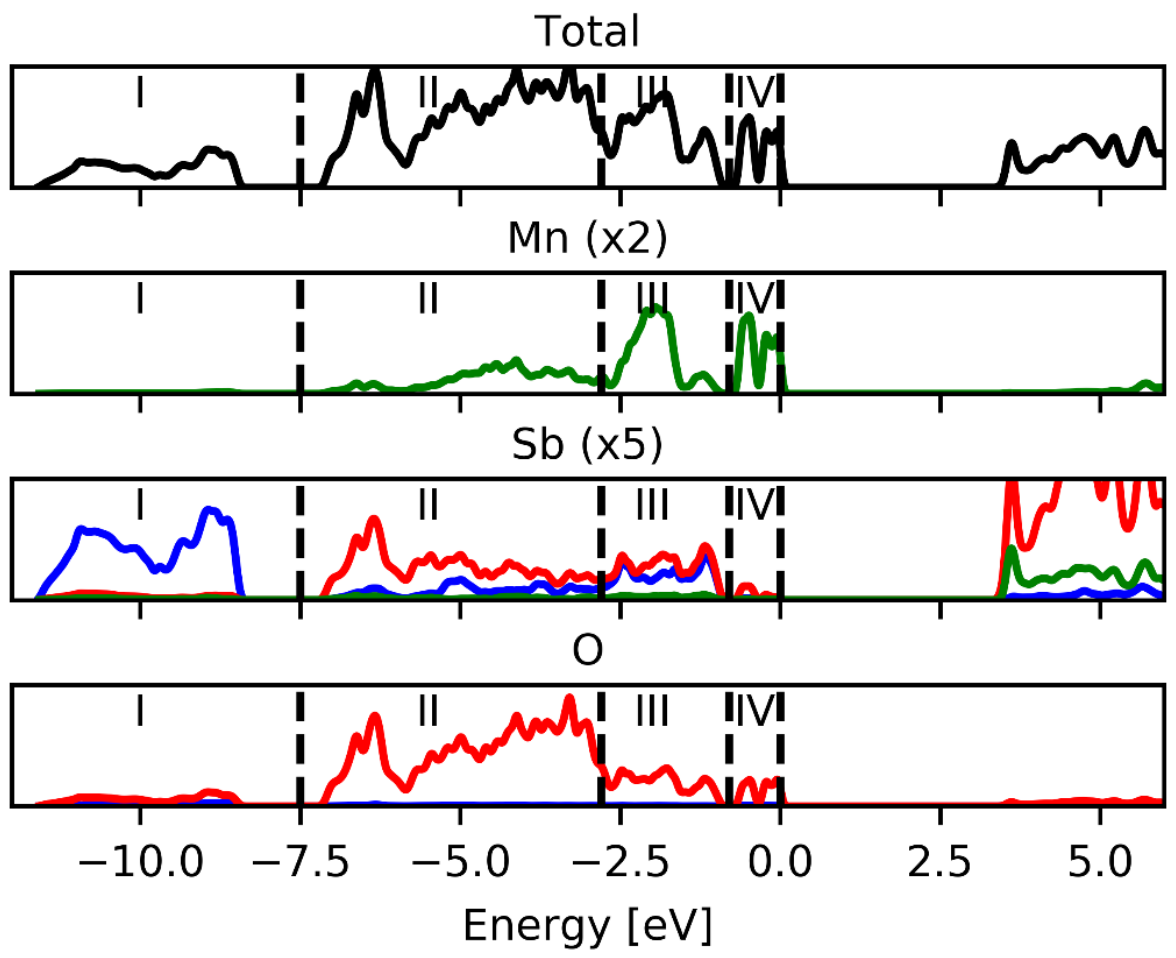

**Figure 2.** Total and orbital projected electronic density of states. Blue lines are s-states, red lines p-states and green lines d-states. Individual atomic contributions have been enlarged to allow visual inspection.

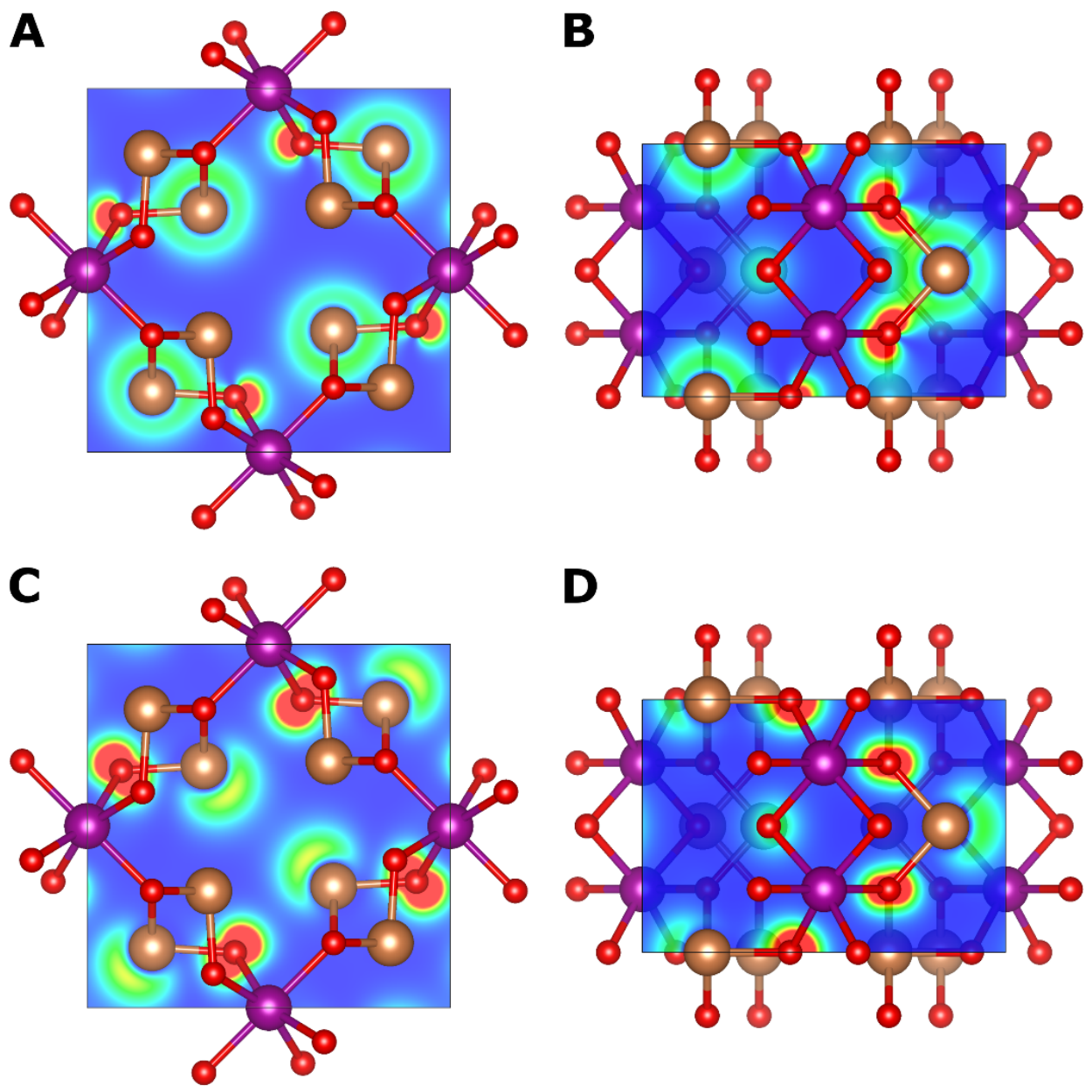

**Figure 3.** Valence electron density in selected planes and regions from Figure 2. (A) and (B) Region I, (C) and (D) Region III. (A) and (C) (001) plane at z=½, (B) and (D) plane spanned by Sb and two equivalent O1. Contours are drawn from 0 (blue) to 0.0445 e $au^{-3}$ (red). The figures were generated using VESTA.[27]

## Real space identification of the lone pairs

So far, we have considered the electronic structure from an orbital projected point of view. However, it is interesting also to analyze the lone pairs and the chemical bonding from the perspective of real space descriptors assessing e.g. the electron localization. A lone pair is characterized by a large degree of electron localization, and it is thus commonly identified using descriptors that assess the localization of electrons or concentration of the electron density. Two commonly used descriptors are, respectively, the electron localization function (ELF)[28-30] and the Laplacian of the electron density, $\nabla^2\rho(\boldsymbol{r})$.[31]

In Figure 4a, the ELF is plotted in the (001) plane at z=½, i.e. at the plane containing the two antimony atoms pointing towards each other. The electrons are seen to be extremely localized at the expected lone pair region on antimony, and furthermore, we see a tendency of the lone pairs to avoid each other, as the maximum localizations are located at a significant angle away from the interatomic line. It is also interesting to note that not only the ELF maxima tend to avoid each other, but the lone pairs are themselves asymmetric. This leads to alignment of high and low electron localizations, which minimizes the unfavorable interactions between the lone pairs.

In Figure 4b, the lone pair is visualized using the Laplacian of the electron density. If plotting it in a 2D plane like the ELF, no obvious features are seen, so instead, we search for critical points in the charge concentration on antimony. Typically, one would perform the search in the outermost valence shell, but for heavy atoms, this is often buried within the charge depletions of the inner shells.[32] Therefore, we use the procedure outlined by Sist *et al.*,[33] who showed that the lone pair character is not only present as a charge concentration in the outermost valence shell, but also in the shell below it. In Figure 4b, the four charge concentrations in the N-shell on Sb are shown, and they correspond well to an $sp^3$-hybridization with three concentrations pointing towards neighboring oxygen atoms, and one charge concentration pointing in the direction where the lone pairs were also observed from the ELF. Also from the charge concentrations, a slight deviation from the interatomic line is observed.

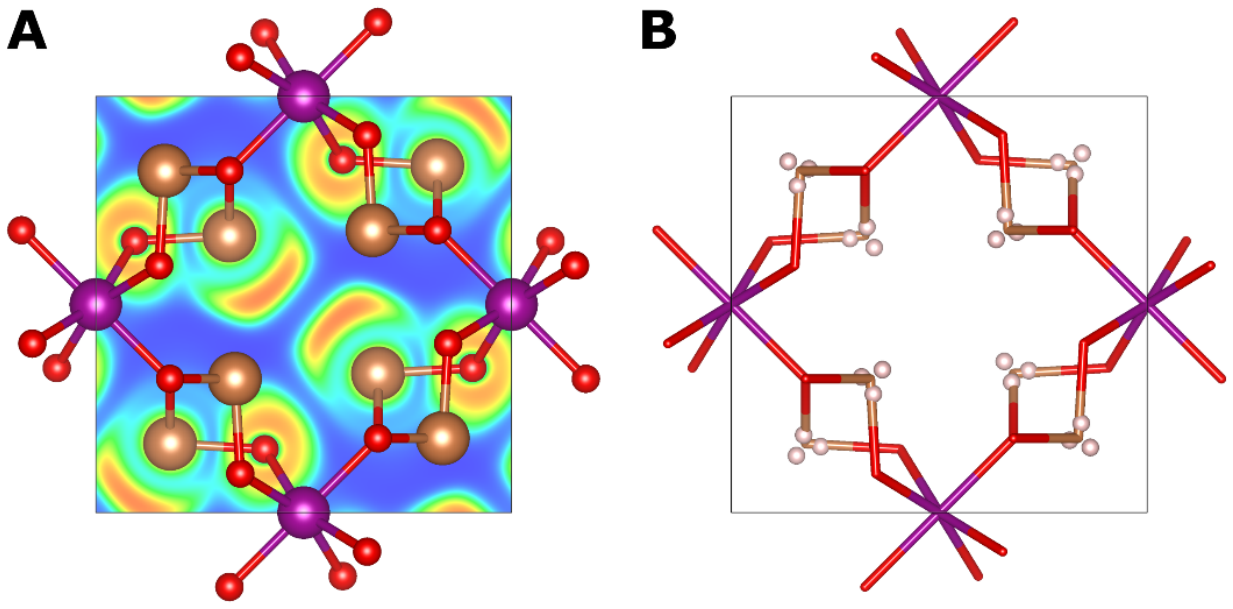

**Figure 4**. (A) ELF in the (001) plane at z=½. Contours are drawn from 0 (blue) to 1 (red). (B) Maxima in $L(\boldsymbol{r}) = -\nabla^2\rho(\boldsymbol{r})$ in the N shell of antimony.

**Pressure effects on the lone pair interaction**

We have now shown that there is a destabilizing interaction due to lone pairs pointing towards each other, which is presumably decreased by introducing an angle between the lone pairs. To further investigate this feature, we simulated the effect of pressure on the structure by optimizing the structure at various constant volumes and extracting the pressure from the energy-volume relation. In Figure 5, the unit cell parameters and bond lengths are shown as a function of pressure. It is seen that the *a*-axis contracts more upon increased pressure, which makes sense since the channels of low electron density run along the *c*-axis, meaning that there is a void space for the structure to relax into in the *ab*-plane. Interestingly, we see that the bond lengths change significantly with pressure, where the originally short Mn-O2 bond stays practically constant with pressure, whereas the Mn-O1 bond length decreases significantly with pressure. Eventually, the order of the bond lengths switches at a very moderate pressure of between 1 and 2 GPa. This is a consequence of the Mn-O1 bond lying in the *ab*-plane, which contracts the most. Here it should be noted that the difference between the Mn-O1 and Mn-O2 bond lengths at zero pressure is smaller in our optimization than from experiment at 100 K, so experimentally the switch may not occur until at a higher pressure (see Table S1 in the SI). For the Sb-O bonds, the changes are less significant, but interestingly enough the short Sb-O2 bond, which lies in the *ab*-plane actually increases slightly as a function of pressure, whereas the longer Sb-O1 bond decreases slightly. One should note that it is assumed that no phase transition occur, which has been observed to occur for the iron analogue, $FeSb_2O_4$, at around 4 GPa.[34]

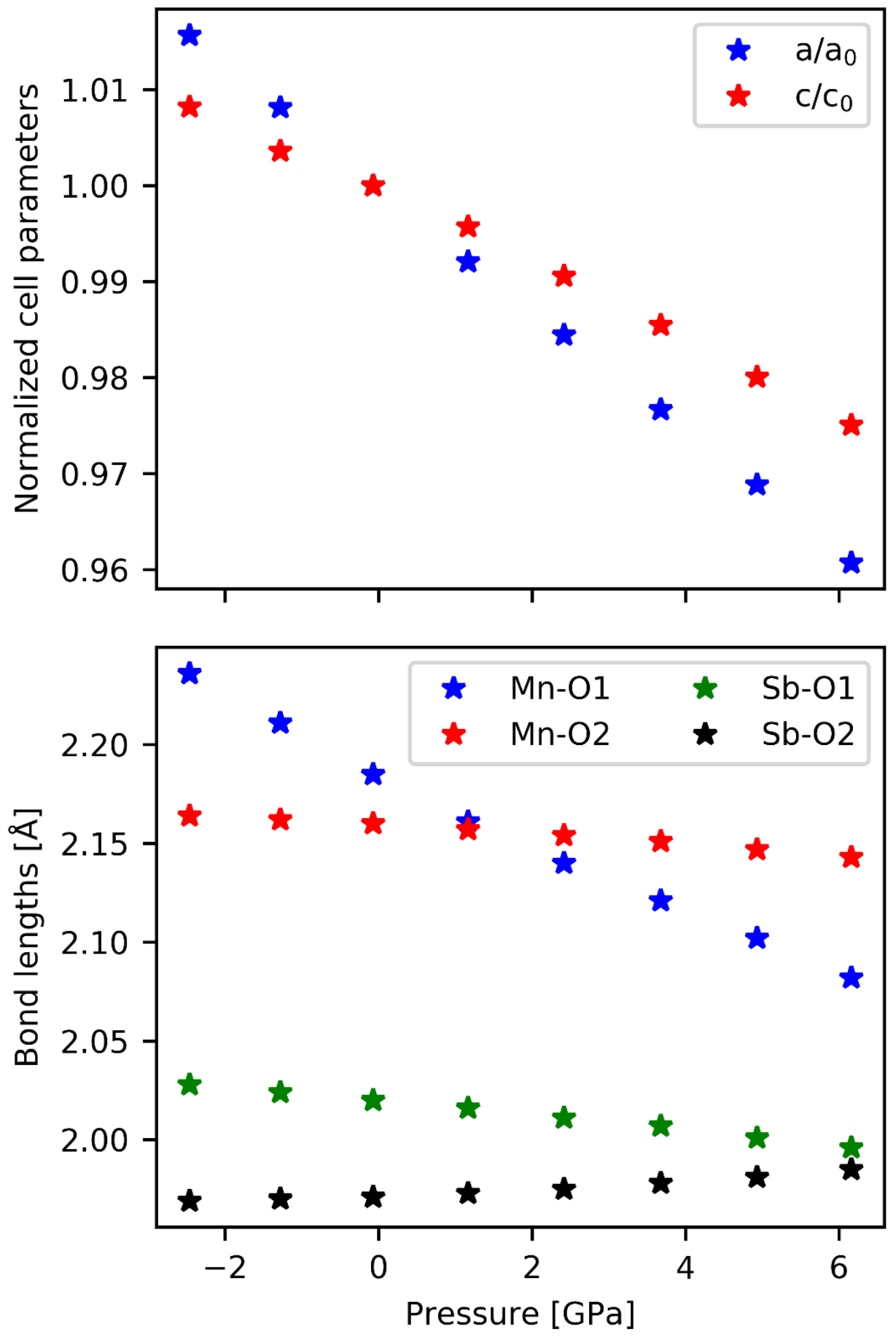

**Figure 5.** Normalized cell parameters and bond lengths of the four unique short bonds as a function of pressure. The optimized geometry used in the previous section is seen to be at slightly low negative pressure from a Birch-Murnaghan third order fit to energy-volume curve.

In Figure 6a, the effect of pressure on the lone pair deflection is shown. It is clearly seen that with decreasing distance between Sb atoms, resulting from the high pressure, the angle between the interatomic vector and the lone pair vector increases. The enhanced angle with increasing pressure (and shorter Sb-Sb distance) is also qualitatively visualized in Figures 6b and 6c, where the ELF is seen at a volume reduced by 6 % and 10 %, respectively, to be compared with the ELF for the optimized volume shown in Figure 4a. This deflection is a clear effect of the repulsion between lone pairs, which use the flexibility of the structure to reduce the repulsion as much as possible.

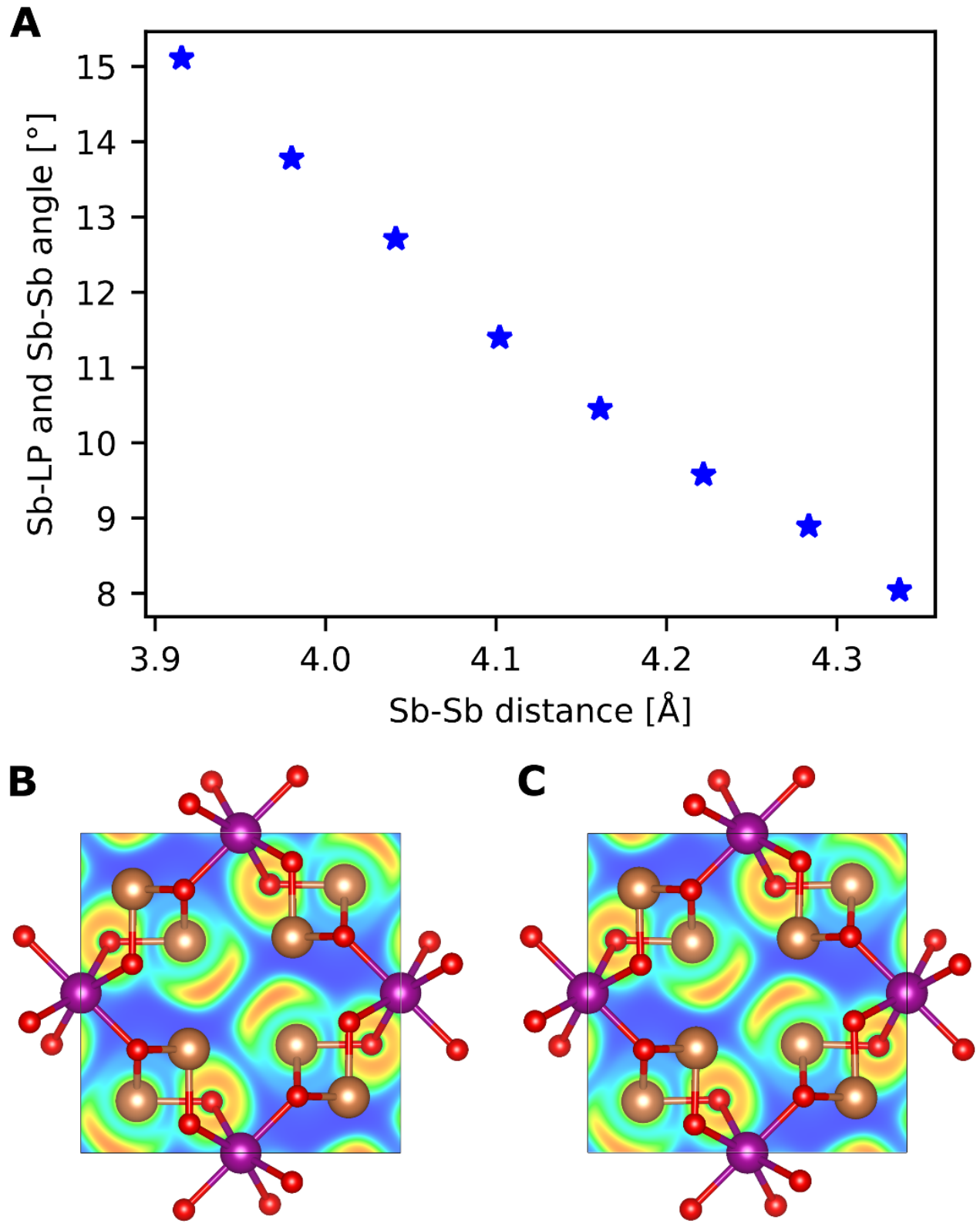

**Figure 6.** (A) Angle between the Sb to lone pair (LP) vector and the Sb to Sb vector as a function of pressure. (B)-(C) ELF in the (001) plane at z=½ for the structure with a volume reduced by 6 % (B) and 10 % (C). Contours are drawn from 0 (blue) to 1 (red).

**Topological analysis of the electron density**

Having analyzed the interactions between the lone pairs on antimony, we now report the general bonding scheme in the material. Since the electron density distribution is a quantum mechanical observable, it is useful to analyze the chemical bonding (almost) purely based on this, which is the foundation of Bader's QTAIM.[24] Here we define the topological atom based on partitioning of the electron density using the zero flux surface, and we define a bonding interaction based on the (3,-1) critical points, i.e. points where the density is minimum along one direction (parallel to the bond) and maximum along two directions (perpendicular to the bond). Based on the properties at the bond critical point (BCP), we can obtain valuable information about chemical bonding based on well-established relations that can often be related to structural and physical properties.[31, 35]

In Table 1, all BCPs are shown. BCPs are found at all four unique bonds drawn in Figure 1, i.e. two unique Mn-O and two unique Sb-O bonds. In addition, BCPs are found between Sb and O2 at larger distance and between the two Sb atoms "through" the lone pair. All bonds have quite low density and positive Laplacian, which is, however, not surprising since the BCPs lie in a region of charge depletion on the heavy atoms and at a distance from the nuclei where the charge distribution of the corresponding isolated atoms is also depleted.[32] Looking at the energy densities, we see that all four short bonds have negative total energy densities, and generally the shorter the bond, the larger the electron density and the more negative the total energy density. Furthermore, the kinetic energy per electron, $G(\boldsymbol{r}_b)/\rho(\boldsymbol{r}_b)$, is slightly above one. All these characteristics are common for polar covalent bonds and donor acceptor bonds. The integrated atomic charges in Table 2 show that large ionic contributions are present in all interactions, since charges are found to be +1.53 and +1.89 on Mn and Sb, respectively. These charges are quite large, showing the large ionic degree of the bonding, although significantly smaller than those corresponding to their formal oxidation states.

In Figure 7, the deformation density and negative Laplacian of the electron density are shown for selected planes. Here, we observe that there is some charge accumulation in all four types of short bonds. It is especially interesting to note that manganese shows positive deformation density towards oxygen, which clearly highlights the covalent contribution to this bond. In the Sb-O bonds, the deformation density is less pronounced and both positive and negative deformation density is observed along the bond path. This is in very good agreement with the observation from both theory and experiment in $Sb_2O_3$.[36] Combined with the fact that charge concentrations on antimony point towards oxygen, this shows that there is some degree of covalency in this bonding, meaning that the Sb-O bonds are of polar covalent type. This is important for reconciling the orbital based view with our analysis of the electron density, since the theory of stereo-chemically active lone pairs requires significant orbital overlap, meaning that there must be a partial covalent character to the bond.

The two extra BCPs found are between atoms at much longer distance, the density is much lower and the total energy density is practically zero. This shows that they are largely present due to the geometry dictated by the stronger interactions, rather than being important structure determining features on their own. Especially the Sb-Sb BCP is clearly not a stabilizing interaction, since we have shown that the charge concentrations try to avoid each other to lower the repulsion between adjacent lone pairs, rather than directing them towards each other as seen in typical covalent bonds.

**Table 1.** Bond critical points and properties evaluated at these points. R is distance between the atoms, $d_1$ and $d_2$ are the distances from the BCP to the first and second atom, respectively, $\rho(r_b)$ and $\nabla^2\rho(r_b)$ are the electron density and Laplacian of the electron density, $G(r_b)$, $V(r_b)$ and $H(r_b)$ are the kinetic, potential and total energy densities at the BCP, and $\varepsilon$ is the ellipticity. The units are as follows (a) Å, (b) eÅ$^{-3}$, (c) eÅ$^{-5}$, (d) Hartree Å$^{-3}$, (e) Hartree e$^{-1}$. The Sb-O2 bond marked by a star is not shown in the structural figures, as it is significantly longer than the typical bond length.

| | **R**[a] | **$d_1$**[a] | **$d_2$**[a] | $\boldsymbol{\rho(r_b)}$[b] | $\boldsymbol{\nabla^2\rho(r_b)}$[c] | $\boldsymbol{G(r_b)}$[d] | $\boldsymbol{V(r_b)}$[d] | $\boldsymbol{H(r_b)}$[d] | $\boldsymbol{G(r_b)/\rho(r_b)}$[e] | $\boldsymbol{\varepsilon}$ |
|---|---|---|---|---|---|---|---|---|---|---|
| Mn-O2 | 2.16 | 1.07 | 1.09 | 0.39 | 6.13 | 0.50 | -0.57 | -0.07 | 1.27 | 0.03 |
| Mn-O1 | 2.18 | 1.08 | 1.10 | 0.35 | 5.69 | 0.45 | -0.50 | -0.05 | 1.26 | 0.04 |
| Sb-O2 | 1.97 | 1.01 | 0.96 | 0.80 | 10.28 | 0.98 | -1.24 | -0.26 | 1.23 | 0.01 |
| Sb-O1 | 2.02 | 1.03 | 0.99 | 0.70 | 9.01 | 0.83 | -1.02 | -0.20 | 1.19 | 0.05 |
| Sb-O2* | 2.84 | 1.49 | 1.35 | 0.14 | 1.33 | 0.09 | -0.09 | 0.00 | 0.67 | 0.01 |
| Sb-Sb | 4.22 | 2.11 | 2.11 | 0.04 | 0.27 | 0.02 | -0.01 | 0.00 | 0.38 | 0.08 |

**Table 2.** Integrated atomic charges (Q) in units of electrons, volumes (V) in $Å^3$ and integrated Lagrangian (L) in atomic units. Only the unique atoms from the structural point of view are shown. The only numerically significant difference between magnetically different atoms are the spin density related properties.

| | **Q** | **V** | **L** |
|---|---|---|---|
| Mn | 1.53 | 9.47 | $-9.1\cdot10^{-3}$ |
| Sb | 1.89 | 23.0 | $2.1\cdot10^{-3}$ |
| O1 | -1.30 | 14.3 | $1.5\cdot10^{-4}$ |
| O2 | -1.35 | 14.3 | $3.3\cdot10^{-4}$ |

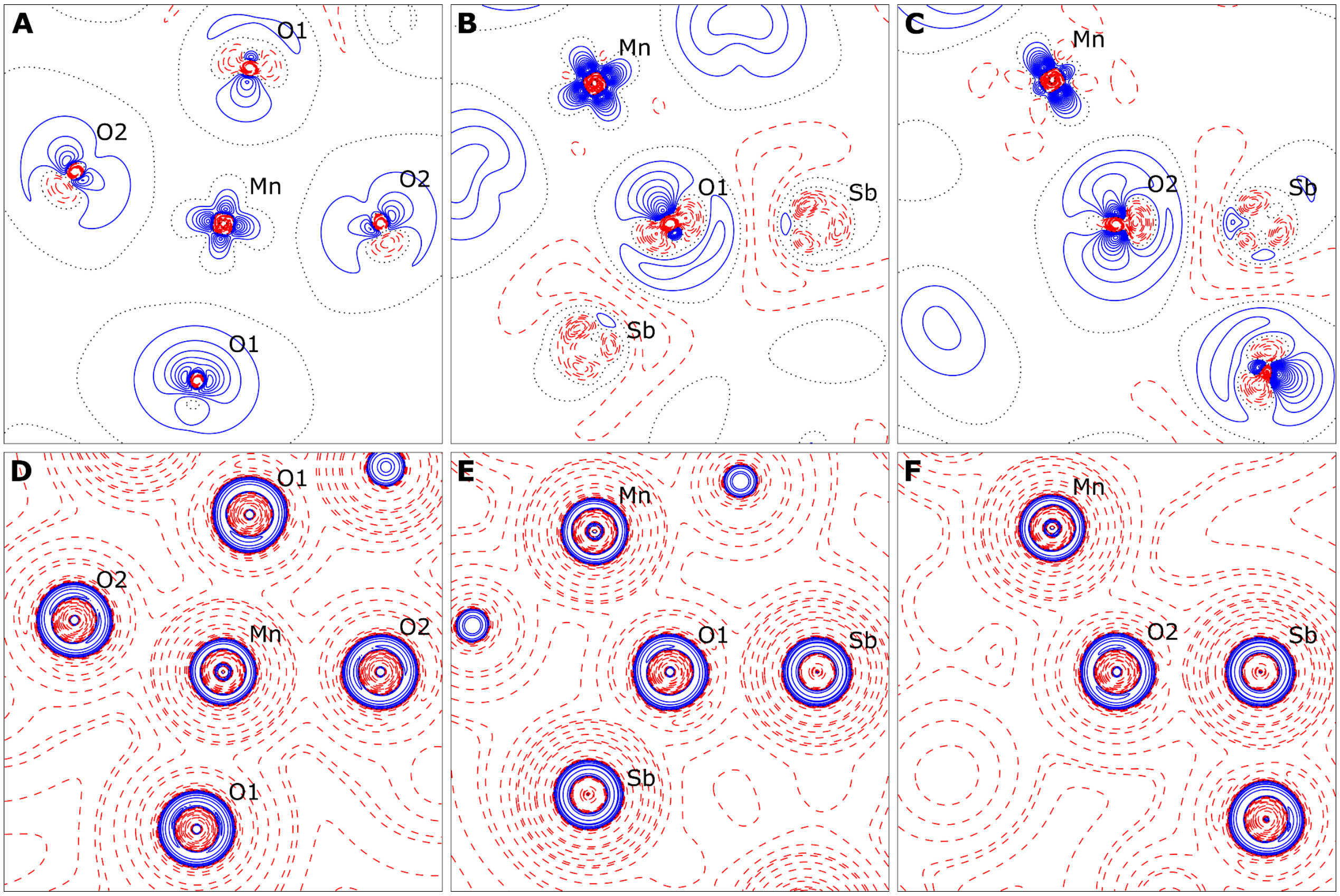

**Figure 7.** (A)-(C) Deformation densities in the planes containing the labelled nuclei, positive (solid blue), negative (dashed red) and zero (dotted black) contours are shown, contour spacings are 0.1 eÅ$^{-3}$ in (A) and 0.05 eÅ$^{-3}$ in (B) and (C). (D)-(F) Negative Laplacian of the electron density in selected planes, positive (solid blue), negative (dashed red) and zero (dotted black) contours are shown, contours are shown at $a \cdot 10^n$ with a = 1, 2, 4, 8 and n = -2, -1, 0, 1, 2, 3, 4.

**Implications for other materials with stereo-chemically active lone pairs**

The clear repulsion between stereo-chemically active lone pairs, which we have demonstrated here, might have important implications for our understanding of crystal structures with expressed stereo-chemically active lone pairs. In almost all other structures, an important structural motif is that lone pairs point into areas where no other lone pairs are present such as in the litharge structure adopted by SnO and PbO, and the GeS structure adopted by GeS, GeSe, SnS and SnSe. Even in a structure like cubic $Sb_2O_3$ (senarmonite), where the lone pair region point towards other Sb atoms, they arrange in a tetrahedron to avoid having charge concentrations pointing directly towards each other.[8] In the present structure, the lone pairs are forced into the same region in pairs, but tend to decrease this unfavorable interaction by introducing a deflection angle. Furthermore, the lone pairs only interact strongly between pairs of atoms, but each of these atomic pairs tend to occupy mutually exclusive regions of space to decrease interactions with other atomic pairs, similarly to the individual behavior in the other mentioned structures.

As mentioned earlier, there are indications that the one unique oxygen atom bonded to two antimony, was much more involved in the lone pair formation than other one, which is only bonded to one Sb atom. This suggests that there could be a cooperative effect involved in the lone pair expression. Skoug and Morelli showed an empirical correlation between bond angles and lattice thermal conductivity in chalcogenides of As, Sb and Bi. In this case, bond angles were used as a measure for the degree of lone pair expression according to valence shell electron pair repulsion (VSEPR) theory, since a more strongly expressed lone pair closer to the nuclei will give rise to smaller bond angles.[11] In their example case of the Cu-Sb-Se system, the main structural difference between the two materials $CuSbSe_2$ and $Cu_3SbSe_3$, where the lone pair is more strongly expressed in the former, is that the $SbSe_3$ units are connected in $CuSbSe_2$, but isolated in $Cu_3SbSe_3$. These two observations motivated us to perform a database search in the Inorganic Crystal Structure Database (ICSD)[37] in order to understand, if it is a general trend that isolated units tend to have larger bond angles arising from weaker lone pair expression, and therefore possibly also lower lattice thermal conductivity. The search included all MSb(III)X structures (with M being an alkali metal, an alkaline-earth metal or a transition metal, and X being O, S or Se) from the ICSD, where the coordination number of antimony is three. Structures with partial occupancies (i.e. disorder) and structures, where the atomic coordinates were not refined, were excluded. The structures were sorted based on whether the $SbX_3$ were isolated or connected. In order to make this distinction, an operational definition of a bond, which only depends on the geometry, must be used. We tried a criterion based on covalent radii, but to include a reasonable number of bonds, an arbitrary increase of the sum of covalent radii must be used. Instead, we chose to use a criterion based on bond valence parameters, $s = \exp[(r_0 - r)/B]$, where $B = 0.37$ is a universal constant, $r_0$ is the tabulated bond valence parameter, $r$ is the bond length and $s$ is the corresponding bond valence.[38, 39] It was suggested by Altermatt and Brown that values larger than 0.6 for the bond valence correspond to a covalent bond, and that values larger than 0.038 times the oxidation state of the cation correspond to a bond to be included in the calculation of the effective valence.[40] However, these two numbers are too strict and too loose, respectively, in the present case for a reasonable number of bonds to be included, so we chose a bond valence value of 0.3 to be the minimum value for a bond to be included in the analysis. This correspond to maximum distances of 2.42 Å, 2.90 Å and 3.02 Å for bonds between Sb and O, S and Se, respectively. Changing the parameters slightly makes a small difference in the final histogram, but does not affect conclusions.

Figure 8 shows that, despite a significant degree of overlap between the two groups, there is a clear tendency for isolated $SbX_3$ units to have larger bond angles than the connected ones. According to VSEPR theory, this means that the lone pair is further away from the nuclei in the isolated cases, or put differently that the lone pair is less expressed. The present results, therefore, suggest that there is a cooperative effect involved in the stronger expression of the lone pair, meaning that the ability of the chalcogen to be involved in lone pair formation becomes stronger, when it is influenced by more than one Sb atom. Since the bond angles, and thus lone pair expression, is strongly correlated with lattice thermal conductivity, this gives an interesting handle for designing new low thermal conductivity structures with applications as thermoelectric materials.

As discussed above, lone pair expression arises from an orbital interaction, i.e. a covalent chemical interaction, so the larger degree of lone pair expression should correspond to a more covalent interaction. Zhang et al.[41] showed that heat basically conducts along covalent bonds in layered systems through a strong correlation between the electron density at the bond critical points and the thermal conductivity. Thus, weaker orbital interaction should lead to lowering of the thermal conductivity. Therefore, an explanation for the lower lattice thermal conductivity in systems with larger bond angles shown by Skoug and Morelli may arise from weaker covalent interactions in these systems, which originates in whether the units are connected or isolated. Indeed, Skoug and Morelli showed that $Cu_3SbSe_3$ with isolated $SbX_3$ units has lower thermal conductivity than $CuSbSe_2$, which has connected units.

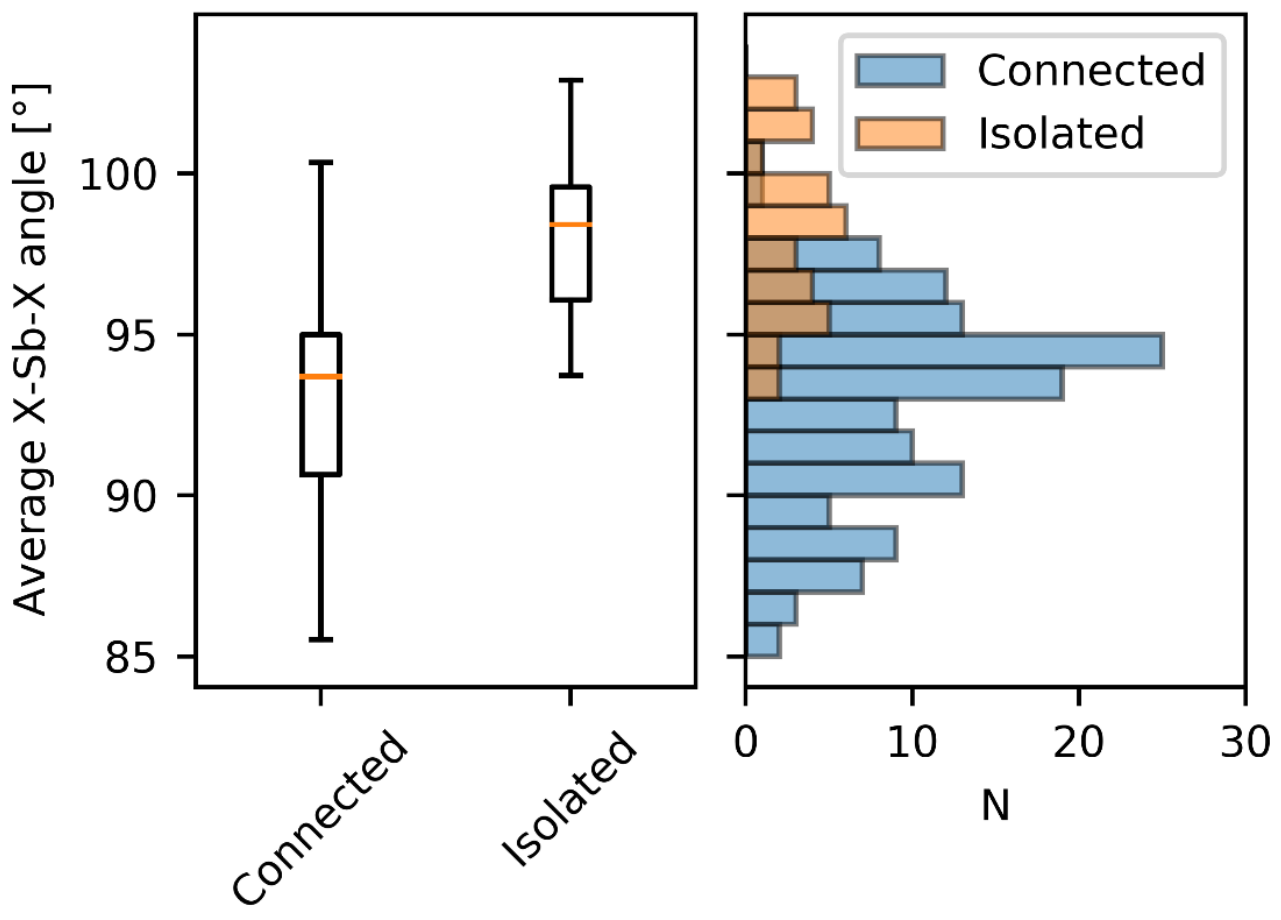

**Figure 8.** Average X-Sb-X for MSb(III)X structures (with M being an alkali metal, an alkaline earth metal or a transition metal, and X being O, S or Se) from the ICSD depending on whether the $SbX_3$ units are connected or isolated. To the left a box plot is shown with the orange bar marking the median value, the box marks the interquartile range and the whiskers mark the spread of the data points. To the right a histogram shows the number of $SbX_3$ units (N) with bond angles within the given interval. Data are binned in 1° intervals. The search is based on 77 crystal structures with 137 unique connected $SbX_3$ units and 35 unique isolated $SbX_3$ units. See Table S2 in the SI for a list of structures and ICSD codes.

Previously, the differences in bond angles have been interpreted as differences in effective atomic valence of the Sb atom.[13] Here a perfect tetrahedron with four neighbors (bond angles of 109.5°) corresponds to an oxidation state of +V and a complete transfer of the 5s lone pair to a bonding interaction with an anion, whereas an oxidation state of +III in a three-fold coordination corresponds to a completely localized lone pair and thus a small bond angle. Intermediate cases then correspond to a progressive change in effective oxidation state and a less localized lone pair, which results in a larger bond angle. This was based on the fact that shorter bond lengths, and thus formally a larger effective valence, were observed for larger bond angles. Alternatively, one can imagine the limit of a flat coordination with 120° angles corresponding to an $sp^2$-hybridized Sb with a lone pair in the remaining p-orbital. Progressive changes towards small bond angles correspond to more s-character in the lone pair, and correspondingly more p-character in the Sb-X bonds, which leads to longer bonds, similar to the increase in the C-H bond lengths with increasing p-character in going from acetylene (sp) through ethylene ($sp^2$) to ethane ($sp^3$),[42] and in agreement with previous observations.[13] Interestingly, it is the smaller bond angles corresponding to the longer bonds that give rise to the largest thermal conductivity. Also in the case of $MnSb_2O_4$, it is in fact the longer Sb-O bond that has the largest degree of covalency determined from the valence electron densities and is more involved in lone pair formation. The present group of materials is thus a counterexample to the otherwise established correlation between shorter bond lengths and higher thermal conductivity, which is present over a very wide range of bond lengths and thermal conductivities, but with several groups of similar systems not following the trend.[12] Thus, it seems to be the covalency of an interaction, rather than the bond length itself, which is important for the thermal conductivity.

The lattice thermal conductivity of a material is determined from several contributions, which can be divided into two groups, those arising from the phonon dispersion itself, e.g. phonon group velocity, and those arising from scattering of phonons, e.g. phonon-phonon scattering from anharmonicity and impurity scattering. Lone pair expression is often discussed in terms of the presence of asymmetric coordination, and the repulsion between the lone pair and valence electrons on neighboring atoms, which is thought to lead to an anharmonic vibration potential, and therefore a decreased thermal conductivity through phonon-phonon scattering.[11, 43]

With the present observations of increased covalency for connected units and correspondence between lone pair expression and covalency, we instead suggest that the lone pair expression and thermal conductivity are mainly related through the effect of vibrations on the electronic structure. In the case of a strongly expressed lone pair and therefore significant covalent interactions, small displacements of the atoms should lead to large changes in total electronic energy, giving large phonon velocities, but only small differences in the spatial structure of the orbitals. For weakly expressed lone pairs, the effect on the total energy should be smaller, giving low phonon velocities, but may give rise to changes in the spatial structure of the orbitals. In this way, a more strongly expressed lone pair and higher degree of covalency can be thought of as corresponding to a more rigid electronic structure. This is most easily understood for cases with coordination number six in perfect octahedral symmetry such as the lead chalcogenides, where the lone pair is not expressed in the average structure, but will be weakly expressed in different spatial regions depending on the direction of vibration, corresponding to a large coupling between atomic displacement and electronic structure. This leads to dynamic local distortions in this type of materials.[44] Similarly, it was recently shown that increased lone pair expression along the Peierls distortion in GeTe is associated with more covalent bonding and a more rigid electronic structure evidenced by a decrease in response properties such as the Born effective charge.[45]

The effect of atomic displacement on the electronic structure may give rise to anharmonic vibrations, e.g. in thermoelectric SnSe where a specific anharmonic mode revealed from inelastic neutron scattering was shown to highly perturb the electronic structure in the regions of interest for the lone pair.[46] However, this is not necessarily the case, and it is important to acknowledge that also the degree of covalency in the system will be important in determining its lattice thermal conductivity. It is clearly of interest to further probe relations between the extent of lone pair expression, thermal conductivity, structural disorder and anharmonic thermal motion.

## Conclusions

In conclusion, we have shown that $MnSb_2O_4$ has all the characteristics of a material with Sb(III) stereo-chemically active lone pairs. We observe significant contribution of antimony 5s and 5p states close to the Fermi level, and from the valence density, this is attributed to an antibonding configuration expressed as a stereo-chemically active lone pair. From real space descriptors such as the ELF and the Laplacian of the electron density, we found the positions of the antimony lone pairs, which were shown to avoid each other inducing an angle between the lone pair and the interatomic line. This deflection was shown to increase with decreasing distance between antimony atoms by simulating the effect of pressure on the material. The chemical bonding in the material was shown to be polar covalent for both antimony-oxygen and manganese-oxygen interactions with large charge transfer contribution, but also significant charge accumulation in the bonding directions. The covalent degree is important for our understanding of the lone pair formation, as the current theory requires a significant degree of orbital overlap. Analysis of the valence electron density suggested that the oxygen atom bonded to two Sb atoms was more involved in lone pair formation than the one bonded to only one Sb atom. This inspired a database search for structures with isolated and connected $SbX_3$ units (where X is a chalcogen), and showed that larger bond angles are generally found for isolated units. These observations suggest a degree of cooperative effect in the lone pair expression. Since heat conduction is normally largest along covalent bond directions, a stronger lone pair expression should lead to a higher thermal conductivity, as is indeed the case for $SbX_3$ structures. This disagrees with suggestions that strong lone pair expression should lead to a lowering of the thermal conductivity through phonon-phonon scattering caused by anharmonic thermal motion due to specific soft potential directions. Rather, it appears that for these chalcogenides, lone pair expression and thermal conductivity may be related through the effects of atomic displacement on the electronic structure.

## Acknowledgement

This work was supported by the Danish National Research foundation (DNRF93) and the Villum Foundation. The theoretical calculations were performed at the Center for Scientific Computing, Aarhus. Affiliation with the Aarhus University Center for Integrated Materials Research (iMAT) is gratefully acknowledged.

## References

1. Snyder, G. J.; Toberer, E. S., Complex thermoelectric materials. *Nat. Mater.* **2008,** *7* (2), 105-114.
2. Zhao, L. D.; Lo, S. H.; Zhang, Y. S.; Sun, H.; Tan, G. J.; Uher, C.; Wolverton, C.; Dravid, V. P.; Kanatzidis, M. G., Ultralow thermal conductivity and high thermoelectric figure of merit in SnSe crystals. *Nature* **2014,** *508* (7496), 373-377.
3. Ramesh, R.; Spaldin, N. A., Multiferroics: progress and prospects in thin films. *Nat. Mater.* **2007,** *6* (1), 21-29.
4. Lencer, D.; Salinga, M.; Grabowski, B.; Hickel, T.; Neugebauer, J.; Wuttig, M., A map for phase-change materials. *Nat. Mater.* **2008,** *7* (12), 972-977.
5. Ogo, Y.; Hiramatsu, H.; Nomura, K.; Yanagi, H.; Kamiya, T.; Hirano, M.; Hosono, H., p-channel thin-film transistor using p-type oxide semiconductor, SnO. *Appl. Phys. Lett.* **2008,** *93* (3), 032113.
6. Zhou, Y.; Wang, L.; Chen, S. Y.; Qin, S. K.; Liu, X. S.; Chen, J.; Xue, D. J.; Luo, M.; Cao, Y. Z.; Cheng, Y. B.; Sargent, E. H.; Tang, J., Thin-film $Sb_2Se_3$ photovoltaics with oriented one-dimensional ribbons and benign grain boundaries. *Nat. Photonics* **2015,** *9* (6), 409-415.
7. Walsh, A.; Watson, G. W., The origin of the stereochemically active Pb(II) lone pair: DFT calculations on PbO and PbS. *J. Solid State Chem.* **2005,** *178* (5), 1422-1428.
8. Walsh, A.; Payne, D. J.; Egdell, R. G.; Watson, G. W., Stereochemistry of post-transition metal oxides: revision of the classical lone pair model. *Chem. Soc. Rev.* **2011,** *40* (9), 4455-4463.
9. Walsh, A.; Watson, G. W., Influence of the anion on lone pair formation in Sn(II) monochalcogenides: A DFT study. *J. Phys. Chem. B* **2005,** *109* (40), 18868-18875.
10. Waghmare, U. V.; Spaldin, N. A.; Kandpal, H. C.; Seshadri, R., First-principles indicators of metallicity and cation off-centricity in the IV-VI rocksalt chalcogenides of divalent Ge, Sn, and Pb. *Phys. Rev. B* **2003,** *67* (12), 125111.
11. Skoug, E. J.; Morelli, D. T., Role of Lone-Pair Electrons in Producing Minimum Thermal Conductivity in Nitrogen-Group Chalcogenide Compounds. *Phys. Rev. Lett.* **2011,** *107* (23), 235901.
12. Zeier, W. G.; Zevalkink, A.; Gibbs, Z. M.; Hautier, G.; Kanatzidis, M. G.; Snyder, G. J., Thinking Like a Chemist: Intuition in Thermoelectric Materials. *Angew. Chem. Int. Ed.* **2016,** *55* (24), 6826-6841.
13. Wang, X. Q.; Liebau, F., Studies on bond and atomic valences. 1. Correlation between bond valence and bond angles in Sb(III) chalcogen compounds: The influence of lone-electron pairs. *Acta Cryst. B* **1996,** *52*, 7-15.
14. Muller-Buschbaum, H., The crystal chemistry of $AM_2O_4$ oxometallates. *J. Alloys Compd.* **2003,** *349* (1-2), 49-104.
15. Roelsgaard, M.; Nørby, P.; Eikeland, E.; Søndergaard, M.; Iversen, B. B., The hydrothermal synthesis, crystal structure and electrochemical properties of $MnSb_2O_4$. *Dalton Trans.* **2016,** *45* (47), 18994-19001.

16. Fjellvag, H.; Kjekshus, A., Crystal and Magnetic Structure of $MnSb_2O_4$. *Acta Chem. Scand. A* **1985,** *39* (6), 389-395.
17. Jibin, A. K.; Reddy, M. V.; Rao, G. V. S.; Varadaraju, U. V.; Chowdari, B. V. R., $Pb_3O_4$ type antimony oxides $MSb_2O_4$ (M = Co, Ni) as anode for Li-ion batteries. *Electrochim. Acta* **2012,** *71*, 227-232.
18. Nørby, P.; Roelsgaard, M.; Søndergaard, M.; Iversen, B. B., Hydrothermal Synthesis of $CoSb_2O_4$: In Situ Powder X-ray Diffraction, Crystal Structure, and Electrochemical Properties. *Cryst. Growth Des.* **2016,** *16* (2), 834-841.
19. Dovesi, R.; Orlando, R.; Erba, A.; Zicovich-Wilson, C. M.; Civalleri, B.; Casassa, S.; Maschio, L.; Ferrabone, M.; De La Pierre, M.; D'Arco, P.; Noel, Y.; Causa, M.; Rerat, M.; Kirtman, B., CRYSTAL14: A Program for the Ab Initio Investigation of Crystalline Solids. *Int. J. Quantum Chem.* **2014,** *114* (19), 1287-1317.
20. Laun, J.; Oliveira, D. V.; Bredow, T., Consistent gaussian basis sets of double- and triple-zeta valence with polarization quality of the fifth period for solid-state calculations. *J. Comput. Chem.* **2018,** *39* (19), 1285-1290.
21. Peintinger, M. F.; Oliveira, D. V.; Bredow, T., Consistent gaussian basis sets of Triple-Zeta valence with polarization quality for solid-State Calculations. *J. Comput. Chem.* **2013,** *34* (6), 451-459.
22. Adamo, C.; Barone, V., Toward reliable density functional methods without adjustable parameters: The PBE0 model. *J. Chem. Phys.* **1999,** *110* (13), 6158-6170.
23. Perdew, J. P.; Burke, K.; Ernzerhof, M., Generalized gradient approximation made simple. *Phys. Rev. Lett.* **1996,** *77* (18), 3865-3868.
24. Bader, R. F. W., *Atoms in Molecules: A Quantum Theory*. Clarendon Press: Oxford, 1990.
25. Gatti, C.; Saunders, V. R.; Roetti, C., Crystal-Field Effects on the Topological Properties of the Electron-Density in Molecular-Crystals - the Case of Urea. *J. Chem. Phys.* **1994,** *101* (12), 10686-10696.
26. Allen, J. P.; Carey, J. J.; Walsh, A.; Scanlon, D. O.; Watson, G. W., Electronic Structures of Antimony Oxides. *J. Phys. Chem. C* **2013,** *117* (28), 14759-14769.
27. Momma, K.; Izumi, F., VESTA: a three-dimensional visualization system for electronic and structural analysis. *J. Appl. Cryst.* **2008,** *41*, 653-658.
28. Becke, A. D.; Edgecombe, K. E., A Simple Measure of Electron Localization in Atomic and Molecular-Systems. *J. Chem. Phys.* **1990,** *92* (9), 5397-5403.
29. Silvi, B.; Savin, A., Classification of Chemical-Bonds Based on Topological Analysis of Electron Localization Functions. *Nature* **1994,** *371* (6499), 683-686.
30. Savin, A.; Nesper, R.; Wengert, S.; Fassler, T. F., ELF: The electron localization function. *Angew. Chem. Int. Ed.* **1997,** *36* (17), 1809-1832.
31. Gatti, C., Chemical bonding in crystals: new directions. *Z. Kristallogr.* **2005,** *220* (5-6), 399-457.
32. Shi, Z.; Boyd, R. J., The Shell Structure of Atoms and the Laplacian of the Charge Density. *J. Chem. Phys.* **1988,** *88* (7), 4375-4377.
33. Sist, M.; Gatti, C.; Nørby, P.; Cenedese, S.; Kasai, H.; Kato, K.; Iversen, B. B., High-Temperature Crystal Structure and Chemical Bonding in Thermoelectric Germanium Selenide (GeSe). *Chem. Eur. J.* **2017,** *23* (28), 6888-6895.
34. Hinrichsen, B.; Dinnebier, R. E.; Rajiv, P.; Hanfland, M.; Grzechnik, A.; Jansen, M., Advances in data reduction of high-pressure x-ray powder diffraction data from two-dimensional detectors: a case study of Schafarzikite ($FeSb_2O_4$). *J. Phys. Condens. Matter* **2006,** *18* (25), S1021-S1037.
35. Tolborg, K.; Iversen, B. B., Electron Density Studies in Materials Research. *Chem. Eur. J.* **2019**.
36. Whitten, A. E.; Dittrich, B.; Spackman, M. A.; Turner, P.; Brown, T. C., Charge density analysis of two polymorphs of antimony(III) oxide. *Dalton Trans.* **2004,** (1), 23-29.

37. Bergerhoff, G.; Hundt, R.; Sievers, R.; Brown, I. D., The Inorganic Crystal-Structure Data-Base. *J. Chem. Inf. Comp. Sci.* **1983,** *23* (2), 66-69.
38. Brown, I. D.; Altermatt, D., Bond-Valence Parameters Obtained from a Systematic Analysis of the Inorganic Crystal-Structure Database. *Acta Cryst. B* **1985,** *41*, 244-247.
39. Brese, N. E.; O'Keeffe, M., Bond-Valence Parameters for Solids. *Acta Cryst. B* **1991,** *47*, 192-197.
40. Altermatt, D.; Brown, I. D., The Automatic Searching for Chemical Bonds in Inorganic Crystal Structures. *Acta Cryst. B* **1985,** *41*, 240-244.
41. Zhang, J. W.; Song, L. R.; Sist, M.; Tolborg, K.; Iversen, B. B., Chemical bonding origin of the unexpected isotropic physical properties in thermoelectric $Mg_3Sb_2$ and related materials. *Nat. Commun.* **2018,** *9*, 4716.
42. Bent, H. A., An Appraisal of Valence-Bond Structures and Hybridization in Compounds of the 1st-Row Elements. *Chem. Rev.* **1961,** *61* (3), 275-311.
43. Zhang, Y. S.; Skoug, E.; Cain, J.; Ozolins, V.; Morelli, D.; Wolverton, C., First-principles description of anomalously low lattice thermal conductivity in thermoelectric Cu-Sb-Se ternary semiconductors. *Phys. Rev. B* **2012,** *85* (5), 054306.
44. Sangiorgio, B.; Bozin, E. S.; Malliakas, C. D.; Fechner, M.; Simonov, A.; Kanatzidis, M. G.; Billinge, S. J. L.; Spaldin, N. A.; Weber, T., Correlated local dipoles in PbTe. *Phys. Rev. Mater.* **2018,** *2* (8), 085402.
45. Raty, J. Y.; Schumacher, M.; Golub, P.; Deringer, V. L.; Gatti, C.; Wuttig, M., A Quantum-Mechanical Map for Bonding and Properties in Solids. *Adv. Mater.* **2019,** *31* (3), 1806280.
46. Li, C. W.; Hong, J.; May, A. F.; Bansal, D.; Chi, S.; Hong, T.; Ehlers, G.; Delaire, O., Orbitally driven giant phonon anharmonicity in SnSe. *Nat. Phys.* **2015,** *11* (12), 1063-1069.

# Supporting information for

## Expression and interactions of stereo-chemically active lone pairs and their relation to structural distortions and thermal conductivity

Kasper Tolborg,[a] Carlo Gatti,[b] Bo B. Iversen[a]*

a) Center for Materials Crystallography, Department of Chemistry and iNANO, Aarhus University, Langelandsgade 140, 8000 Aarhus C, Denmark

b) CNR-SCITEC Istituto di Scienze e Tecnologie Chimiche “Giulio Natta”, via Golgi section, via Golgi 19, 20133 Milano, Italy

*Corresponding author: Bo B. Iversen, bo@chem.au.dk

**Supporting Note 1: Geometry optimization**

The full geometry, i.e. unit cell and fractional coordinates, were optimized starting from the structure from Roelsgaard et al.[1] according to the information given in the manuscript. In Table S1, a comparison of the optimized structure with the 100 K experimental structure is given.

**Table S1.** Optimized geometry and experimental geometry. The numbers in parentheses are the percentage deviation from experimental values

| Parameter | Optimized | Experimental |
|---|---|---|
| a (Å) | 8.65577671 (-0.40 %) | 8.6905 |
| c (Å) | 6.01823632 (+0.26 %) | 6.0024 |
| Mn | (0,0.5,0.25) | (0,0.5,0.25) |
| Sb | (0.17886,0.16577,0.5) | (0.17912,0.16801,0.5) |
| O1 | (0.17847,0.32153,0.25) | (0.1795,0.3205,0.25) |
| O2 | (0.40618,0.15253,0.5) | (0.4005,0.1450,0.5) |
| Mn-O1 (Å) | 2.1847 (-0.83 %) | 2.203 |
| Mn-O2 (Å) | 2.1602 (+0.85 %) | 2.142 |
| Sb-O1 (Å) | 2.0203 (+0.94 %) | 2.002 |
| Sb-O2 (Å) | 1.9710 (+1.91 %) | 1.934 |

**Supporting note 2: ICSD search**

**Table S2.** All structures with space group number and ICSD code used in the final analysis of bond angles in $SbX_3$ (X=O,S,Se) units. There are 77 structures in total with 137 unique connected $SbX_3$ units and 37 unique isolated $SbX_3$ units.s

| Chemical Formula | Space Group number | ICSD code | Connected units | Isolated units |
|---|---|---|---|---|
| $MgSb_2O_4$ | 135 | 4122 | 1 | 0 |
| $Ba_3(SbO_3)_2$ | 2 | 413764 | 0 | 3 |
| $Sr_5Sb_{22}O_{38}$ | 14 | 425765 | 5 | 0 |
| $Na_3(SbO_3)$ | 217 | 23346 | 0 | 1 |
| $LiSbO_2$ | 14 | 262075 | 2 | 0 |
| $K_3(SbO_3)$ | 198 | 279579 | 0 | 1 |
| $Cs_3(SbO_3)$ | 198 | 279580 | 0 | 1 |
| $KSb_3O_5$ | 14 | 28493 | 2 | 0 |
| $NaSb_5O_8$ | 2 | 28494 | 3 | 0 |
| $Cs_2SbO_{2.5}$ | 8 | 411212 | 2 | 0 |
| $Rb_2(Sb_8O_{13})$ | 11 | 412329 | 7 | 0 |
| $Fe(Sb_2O_4)$ | 14 | 155152 | 2 | 0 |
| $(Sb_4O_5)(ReO_4)_2$ | 2 | 165619 | 4 | 0 |
| $Cu(Sb_2O_4)$ | 106 | 190731 | 1 | 0 |
| $Sb_2VO_5$ | 62 | 203130 | 1 | 0 |
| $MnSb_2O_4$ | 135 | 243444 | 1 | 0 |
| $CoSb_2O_4$ | 135 | 262627 | 1 | 0 |
| $Sb_2VO_5$ | 15 | 27800 | 1 | 0 |
| $ZnSb_2O_4$ | 135 | 31996 | 1 | 0 |
| $FeO(Sb_2O_3)$ | 135 | 4459 | 1 | 0 |
| $Sb_2(MoO_6)$ | 2 | 59292 | 8 | 0 |
| $Sb_2WO_6$ | 1 | 75595 | 2 | 0 |
| $NiSb_2O_4$ | 135 | 86492 | 1 | 0 |
| $Ba_4Sb_4Se_{11}$ | 58 | 31500 | 3 | 1 |
| $BaSb_2Se_4$ | 14 | 32040 | 2 | 0 |
| $Ba_3Sb_2Se_7$ | 15 | 429279 | 0 | 2 |
| $KSbSe_2$ | 2 | 100125 | 1 | 0 |
| $CsSbSe_2$ | 14 | 20773 | 2 | 0 |
| $K_2(Sb_4Se_8)$ | 2 | 402886 | 1 | 0 |
| $Rb_2(Sb_4Se_8)$ | 2 | 402887 | 1 | 0 |
| $Na_3SbSe_3$ | 198 | 425125 | 0 | 1 |
| $CsSb_2Se_4$ | 2 | 61220 | 1 | 0 |
| $RbSb_3Se_5$ | 14 | 64672 | 3 | 0 |
| $KSbSe_2$ | 1 | 660008 | 2 | 0 |
| $K_3SbSe_3$ | 198 | 89607 | 0 | 1 |
| $Rb_3SbSe_3$ | 198 | 89608 | 0 | 1 |
| $Cs_3SbSe_3$ | 198 | 89609 | 0 | 1 |
| $CuSbSe_2$ | 62 | 238476 | 1 | 0 |
| $Cu_3SbSe_3$ | 62 | 401095 | 0 | 1 |
| $Ag_5SbSe_4$ | 36 | 427307 | 0 | 1 |
| $Ag_3SbSe_3$ | 62 | 427308 | 0 | 3 |
| $SbCrSe_3$ | 62 | 84866 | 1 | 0 |
| $Ca_2Sb_2S_5$ | 14 | 201044 | 1 | 1 |
| $Sr_3Sb_4S_9$ | 33 | 201400 | 3 | 0 |
| $Ba_8(Sb_6S_{17})$ | 13 | 26434 | 2 | 3 |
| $BaSb_2S_4$ | 14 | 38372 | 2 | 0 |
| $Ba_3Sb_2S_7$ | 15 | 429278 | 0 | 2 |
| $Sr_6Sb_6S_{17}$ | 19 | 90094 | 6 | 0 |
| $RbSbS_2$ | 1 | 200263 | 2 | 0 |
| $CsSbS_2$ | 14 | 200798 | 1 | 0 |
| $Cs_4Sb_{14}S_{23}$ | 2 | 20866 | 14 | 0 |
| $Cs_2Sb_4S_7$ | 14 | 2193 | 3 | 0 |
| $Rb_2Sb_4S_7$ | 2 | 2194 | 2 | 0 |
| $K_2Sb_4S_7$ | 15 | 25329 | 1 | 0 |
| $K(Sb_5S_8)$ | 7 | 410178 | 7 | 0 |
| $Li_3SbS_3$ | 33 | 424834 | 0 | 1 |
| $Na_3SbS_3$ | 198 | 425458 | 0 | 1 |
| $Cs_3SbS_3$ | 198 | 426551 | 0 | 1 |
| $K_3SbS_3$ | 198 | 426552 | 0 | 1 |
| $Rb_3SbS_3$ | 198 | 426554 | 0 | 1 |
| $Li_3S_{18}Sb_{11}$ | 2 | 433665 | 10 | 0 |
| $RbSbS_2$ | 2 | 56788 | 1 | 0 |
| $Cs_2(S_2)(Sb_4S_6)$ | 2 | 67976 | 1 | 0 |
| $AgSbS_2$ | 9 | 16578 | 2 | 0 |
| $CuSbS_2$ | 62 | 171051 | 1 | 0 |
| $HgSb_4S_8$ | 15 | 174377 | 4 | 0 |
| $Ag_3SbS_3$ | 161 | 181518 | 0 | 1 |
| $Ag_5SbS_4$ | 36 | 290662 | 0 | 1 |
| $Ag_3(SbS_3)$ | 14 | 33714 | 0 | 1 |
| $Cu_3(SbS_3)$ | 19 | 403113 | 0 | 1 |
| $MnSb_2S_4$ | 12 | 411178 | 2 | 0 |
| $MnSb_2S_4$ | 62 | 56379 | 2 | 0 |
| $CrSbS_3$ | 62 | 74601 | 1 | 0 |
| $Cu_3SbS_3$ | 14 | 74901 | 0 | 2 |
| $AgSbS_2$ | 5 | 85130 | 2 | 0 |
| $FeSb_2S_4$ | 62 | 93911 | 2 | 0 |
| $AgSbS_2$ | 15 | 94647 | 1 | 0 |

**Supplementary references**

1. Roelsgaard, M.; Nørby, P.; Eikeland, E.; Søndergaard, M.; Iversen, B. B., The hydrothermal synthesis, crystal structure and electrochemical properties of $MnSb_2O_4$. *Dalton Trans.* **2016,** *45* (47), 18994-19001.